\pdfoutput=1 
\documentclass{llncs}
\usepackage{color,graphicx}
\graphicspath{{pics/}{./}}
\usepackage{textcomp}
\usepackage{enumerate,paralist,subfigure,multicol}

\usepackage[a-3u]{pdfx}
\usepackage[timezone=+10]{attachfile2}
\def\addtodocassocfiles#1{%
 \xdef\docassociatedfiles{\docassociatedfiles\space#1 0 R}}%
\def\linkdocassocfiles{%
  \pdfcatalog{/AF[\docassociatedfiles]}%
}
\def\docassociatedfiles{}
\AtEndDocument{\linkdocassocfiles}

\usepackage[noTeX]{mmap}
\hypersetup{unicode,bookmarksopen,%bookmarksnumbered,  
  colorlinks, linkcolor=blue, citecolor=blue, %pagecolor=white,
  urlcolor=blue,filecolor=magenta,
  breaklinks,bookmarksdepth=2}

\providecommand{\AtEndAfterFileList}[1]{}

\DeclareRobustCommand{\dblquote}[1]{{\textquotedblleft#1\textquotedblright}}

\DeclareRobustCommand{\subfiglabel}[1]{\llap{\hypertarget{fig#1}{ }\kern15pt}}
\DeclareRobustCommand{\subfigref}[2][]{\hyperlink{fig#2}{\ifx\relax#1\relax#2\else#1\fi}}

\makeatletter
\DeclareRobustCommand{\La}{L\kern-.36em%
        {\sbox\z@ T%
         \vbox to\ht\z@{\hbox{\check@mathfonts
                              \fontsize\sf@size\z@
                              \math@fontsfalse\selectfont
                              A}%
                        \vss}%
        }%
        \kern-.15em%
}

\DeclareRobustCommand\SMC{%
  \ifx\@currsize\normalsize\small\else
   \ifx\@currsize\small\footnotesize\else
    \ifx\@currsize\footnotesize\scriptsize\else
     \ifx\@currsize\large\normalsize\else
      \ifx\@currsize\Large\large\else
       \ifx\@currsize\LARGE\Large\else
        \ifx\@currsize\scriptsize\tiny\else
         \ifx\@currsize\tiny\tiny\else
          \ifx\@currsize\huge\LARGE\else
           \ifx\@currsize\Huge\huge\else
            \small\SMC@unknown@warning
 \fi\fi\fi\fi\fi\fi\fi\fi\fi\fi
}
\providecommand\SMC@unknown@warning{%
  \typeout{\string\SMC: nonstandard
    text font size command -- using \string\small}}
\makeatother

\providecommand\acro[1]{\textSMC{#1}\@}
\providecommand\textSMC[1]{{\SMC #1}}
\providecommand{\pdt}[1]{\textsf{#1}}% PDF tag name
\newcommand{\BNF}[2][]{\langle\mathit{#2}#1\rangle}

\def\Exfootmark{\hyperlink{Hfootnote.1}{\footnotemark[1]}}
\def\Octalfootmark{footnote \hyperlink{Hfootnote.11}{11}}

\DeclareRobustCommand\PDF{\acro{PDF}}

\def\PDFA{\acro{PDF/A}}
\def\PDFUA{\acro{PDF/UA}}
\def\ISO{\acro{ISO}}
\DeclareRobustCommand\MathML{Math\acro{ML}}

\def\LNAIlink{http://www.springer.com/computer/theoretical+computer+science/book/978-3-319-08433-6?otherVersion=978-3-319-08433-6}

\title{PDF/A-3u as an archival format\\ for Accessible mathematics\thanks
 {The final publication is available at \protect\href{http://link.springer.com/search?query=LNAI+8543}{http://link.springer.com/}, as: \protect\newline
  S.M.~Watt et al. (Eds.): CICM\,2014, \protect\href{\LNAIlink}{LNAI\,8543}, pp.\,184--199, 2014.%\protect\newline
  }}
\titlerunning{PDF/A-3u for Accessible mathematics}
\toctitle{PDF/A-3u as an archival format for Accessible mathematics}
\author{Ross Moore}
\institute{Macquarie University, Sydney, Australia\\
 \email{ross.moore@mq.edu.au}%,\\WWW home page:  \rosshomepage
}
\begin{document}
\maketitle

\begin{abstract}
Including \LaTeX\ source of mathematical expressions, within the PDF document
of a text-book or research paper, has definite benefits regarding `Accessibility' considerations.
Here we describe three ways in which this can be done, fully compatibly
with international standards  ISO\,32000, % {\ISO} 
ISO\,19005-3, % \cite{pdfA3},
and the forthcoming ISO\,32000-2 (PDF\ 2.0). 
Two methods use embedded files, also known as `attachments', 
holding information in either \LaTeX\ or \MathML\ formats, 
but use different PDF structures to relate these attachments to regions of the
document window. % where the visual presentation can be seen. 
One uses structure, so is applicable to a fully `Tagged PDF' context,
while the other uses \pdt{/AF} tagging of the relevant content.
The third method requires no tagging at all, instead including the source
coding as the \pdt{/ActualText} replacement of a so-called `fake space'. 
Information provided this way is extracted via simple \pdt{Select}/\pdt{Copy}/\pdt{Paste} actions, 
and is available to existing screen-reading software and assistive technologies.
\end{abstract}

\section{Introduction}
{PDF/A} is being adopted by publishers and Government agencies for the long-term
preservation of important documents in electronic form. 
There are a few variants, which pay more or less regard to Accessibility considerations; 
i.e., `a' for \emph{accessible}, `b' for \emph{basic}, `u' for (presence of) \emph{unicode mappings} for all font characters. 
Later versions \cite{pdfA2,pdfA3} of this {\ISO} standard \cite{pdfA} 
allow for other file attachments in various data formats.
In particular, the {PDF/A-3u} variant allows the inclusion of \emph{embedded files} of arbitrary types, 
to convey supplementary descriptions of technical portions of a document's contents.

`Accessibility' is more relevant for reports and text-books than for research outputs. 
%Regarding `Accessibility', 
In fact in some countries it is a legal requirement that when a visually-impaired student 
enrols in unit of study for which a text-book is mandated as `Required', 
then a fully accessible version of the contents of that book \emph{must} be made available. % to the student. 
Anecdotally, visually-impaired students of mathematics and related fields much prefer mathematical
material to be made available as \LaTeX\ source, to any other format.
With a Braille reader, this is text-based and sufficiently compact that expressions %mathematical content 
can be read and re-read with ease, until a full understanding has been achieved.
This is often %far 
preferable to having an audio version \cite{raman1,raman2}, 
which is %typically 
less-easy to navigate. 
Of course having \emph{both} a well-structured audio version, as well as textual source,
%would be 
is even more useful. 
The PDF example \cite{ExPDFUA} accompanying this paper\footnote{%
\dots\ should be attached prior to the `References', else downloadable online; see~\protect\cite{ExPDFUA}.}
in fact has both, 
though here we concentrate on how the latter is achievable within PDF documents. 

Again anecdotally, the cost of reverse-engineering\footnote{%
\dots\ with prior permission granted by the publisher \dots}
all the mathematical expressions within a complete textbook
is typically of the order of \pounds10,000 or AUD\,30,000 or CAD\,10,000.
This cost would have been dramatically reduced if the PDF had originally been created 
to include a \LaTeX\ or \MathML\ description of each expression\footnote{%
This is distinct from including the complete \LaTeX\ source of the whole document.
There are many reasons why an author, and hence the publisher,
might not wish to share his/her manuscript;
perhaps due to extra information commented-out throughout the source,
not intended for general consumption.},
attached or embedded for recovery by the PDF reader or other assistive technology. 
How to do this in PDF is the purpose of this paper.

The method of \emph{Associated Files}, which is already part of  the {PDF/A-3} standard \cite{pdfA3}, 
is set to also become part of the {\ISO} 32000-2 (PDF\,2.0) standard \cite{PDF20},
which should appear some time in 2014 or 2015.
In Sect.\,\ref{sect2} this mechanism is discussed in more detail, 
showing firstly how to include the relevant information as attachments, 
which can be extracted using tools in the PDF browser.
The second aspect is to relate the attachments to the portion of content as seen onscreen,
or within an extractable text-stream.
This can be specified conveniently in two different ways. 
One way requires structure tagging to be present (i.e., a `Tagged PDF' document), 
%but 
while the other uses %does not, instead using
direct tagging with an \pdt{/AF} key within the content stream.
In either case a PDF reader needs to be aware of the significance of this \pdt{/AF} key and its
associated %attachments. 
embedded files.

With careful use of the \pdt{/ActualText} attribute of tagged content,
\LaTeX\ (or other) source coding of mathematical expressions can be included within a PDF document, 
virtually invisibly, yet extractable using normal \pdt{Select}/\pdt{Copy}/\pdt{Paste} actions.
A~mechanism, using very small space characters inserted before and after each mathematical
expression, is discussed in Sect.\,\ref{sect3}.
This is applicable with \emph{any} PDF file, not necessarily PDF/A.
It is important that these spaces not interfere with the high-quality layout of the visual content in the document, 
so we refer to them as `fake spaces'.

The various Figures in this paper illustrate the ideas and provide a look 
at the source coding of a PDF document\Exfootmark\ that includes all the stated methods, 
thus including the \LaTeX\ source of each piece of mathematical content. 
(Where explicit PDF coding is shown, the whitespace may have been massaged 
to conserve space within the pages of this paper.)
Indeed the example document includes as many as 7~different representations of each piece 
of mathematical  content:\par
\vspace{-.5\baselineskip}%
\begin{itemize}
\item
 the visual form, as typically found in a PDF document;
\item
 the \LaTeX\ source, in two different ways; % as described above;
  i.e, an attachment associated with a~\pdt{/Formula} structure tag 
  and also associated directly to the (visual) content,
  and as the \pdt{/ActualText} replacement of a `fake space'.
\item
 a \MathML\ version as an attachment, also associated to the \pdt{/Formula} structure tag
  and also associated directly to the (visual) content;
\item
 a \MathML\ representation through the structure tagging;
\item
 words for a phonetic audio rendering, to be spoken by `\textsf{Read Out Loud}';
\item
 the original \LaTeX\ source of the complete document, 
 as a file attachment associated with the document as a whole.
\end{itemize}
In practice not all these views need be included to satisfy `Accessibility' or other requirements.
But with such an array of representations, it is up to the PDF reading software 
to choose those which it wants to support, or which to extract according to particular
requirements of end-users. %, or in response to an explicit request. 
It is remarkable that a single document can be so enriched,
yet still be conforming with a standard such as PDF/A-3u, see \cite{pdfA3}.
Indeed, with all content being fully tagged, 
this document\Exfootmark\ would also validate for the stricter PDF/A-3a standard, 
apart from the lack of a way to specify the proper r\^ole of \MathML\ structure tagging, 
so that tags and their attributes are preserved under the `\textsf{Save As Other ... XML 1.0}' 
export method when using Adobe's `Acrobat~Pro' software.
This deficiency will be %expected to be 
addressed in PDF\,2.0 \cite{PDF20}.

%The %basis for the 
Methods used to achieve the structure tagging in the example document\Exfootmark\ 
have been the subject of previous talks and papers \cite{DML2009,CICM2013} by the author.
It is not the intention here to promote those methods, but rather to present the possibilities
for mathematical publishing and `Accessibility' that have been opened up by the PDF/A-3
and PDF/UA standards \cite{pdfA3,PDF-UA1}, and the `fake spaces' idea.
The example document \cite{ExPDFUA} is then just a `proof-of-concept' to illustrate these possibilities.

Since the PDF/A-3 standard \cite{pdfA3} %these standards are 
is so recent, and with PDF\,2.0 \cite{PDF20} yet to emerge,
software is not yet available that best implements %makes best use of 
the `Associated Files' concept.
The technical content of % portions of 
the Figures is thus intended to assist %enable %for 
PDF software developers in building %to build 
better tools in support of accessible mathematics. 
It %They illustrates 
details \begin{inparaenum}[(i)]
\item
exactly what kind of information needs to be included;
\item
the kind of structures that need to be employed; and
\item
how the information and structures relate to each other.
\end{inparaenum}
For those less familiar with PDF coding, the source snippets have been annotated with high-lighting\footnote{%
\dots\ with consistent use of colours, in the PDF version of this paper \dots } %
and extra words %notations 
indicating the ideas and intentions captured within each PDF object.
Lines are used to show %direct 
relationships between objects %present 
within the same Figure,
or `\textsf{see Fig. Xx}' is used where the relationship extends to parts of coding shown within %explicitly within 
%part of 
a different Figure.
Section~\ref{pdf-overview}
is supplied to give an overview of the PDF file structure and %important 
language features 
so that the full details in the Figures can be better understood and their r\^ole appreciated.

\section{Overview of the PDF file format}\label{pdf-overview}%
PDF files normally come employing %having employed 
a certain amount of compression, to reduce file-size,
so appear to be totally intractable to reading by a human.
Software techniques exist to undo the compression, or the PDF file may have been created without using any.
The %attached
example document\Exfootmark\  %\cite{ExPDFUA} 
was created without compression,
so can be opened for reading in most editing software.

\begin{figure}[pth]
%   Please do not adjust anything that affects vertical spacing within this Figure.
%   The subfigure captions have been carefully positioned over the large image.
\subfiglabel{1a}%
\vspace{-12pt}%
\setbox0=\hbox{%
 \includegraphics[]{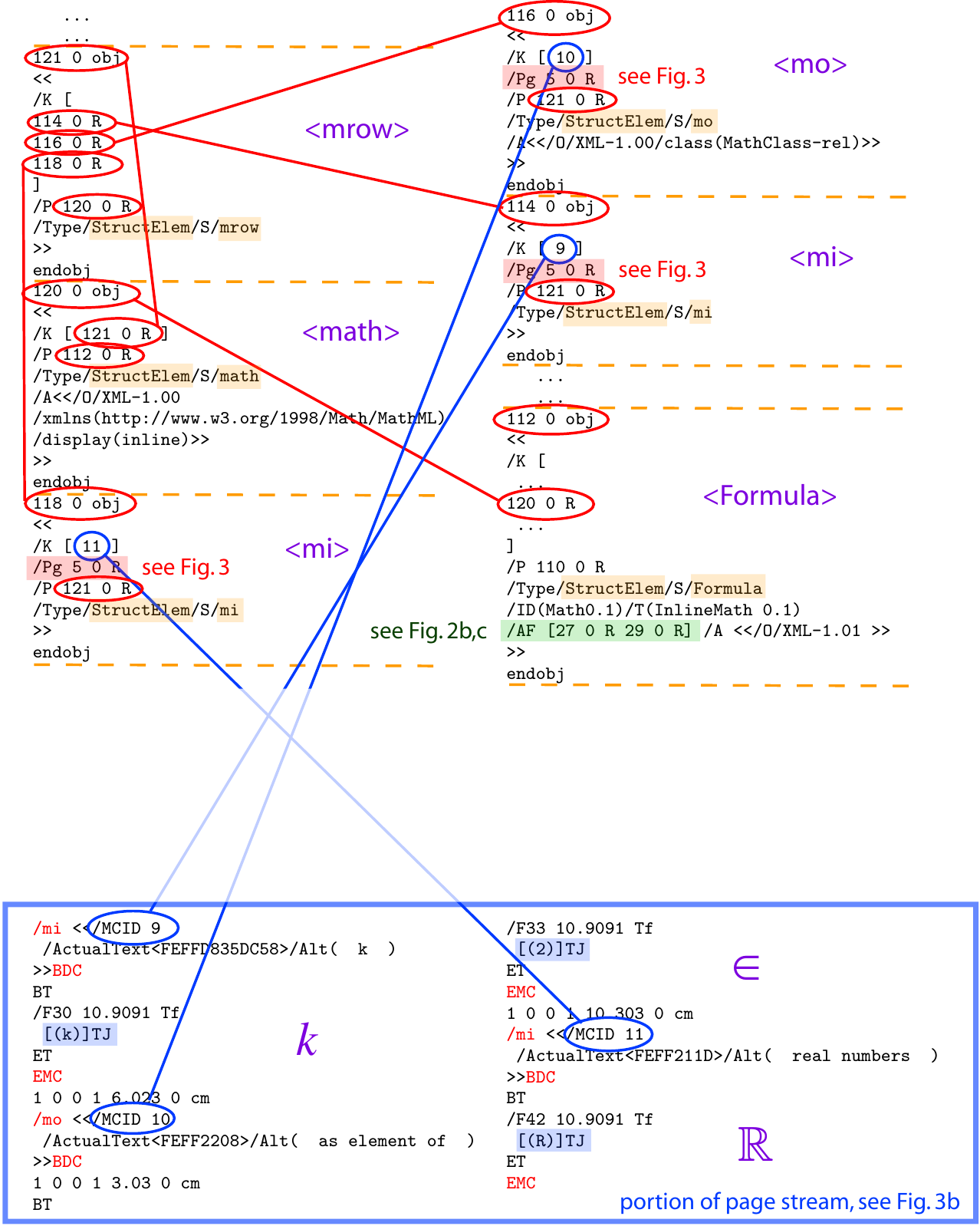}}%-opt-2}}% don't use the 'optimized' version; some stream errors
\setbox0=\hbox{\kern-10pt\lower\ht0\box0}%
\ht0=0pt \dp0=0pt \box0\par
\setbox0=\vbox{\hsize=\linewidth
{\scriptsize
\begin{multicols}2
\begin{verbatim}
   ...
   ...
121 0 obj
<<
/K [
114 0 R
116 0 R
118 0 R
]
/P 120 0 R
/Type/StructElem/S/mrow
>>
endobj
120 0 obj
<<
/K [ 121 0 R ]
/P 112 0 R
/Type/StructElem/S/math 
/A<</O/XML-1.00
/xmlns(http://www.w3.org/1998/Math/MathML)
/display(inline)>>
>>
endobj
118 0 obj
<<
/K [ 11 ]
/Pg 5 0 R
/P 121 0 R
/Type/StructElem/S/mi
>>
endobj
\end{verbatim}
\begin{verbatim}
116 0 obj
<<
/K [ 10 ]
/Pg 5 0 R
/P 121 0 R
/Type/StructElem/S/mo 
/A<</O/XML-1.00/class(MathClass-rel)>>
>>
endobj
114 0 obj
<<
/K [ 9 ]
/Pg 5 0 R
/P 121 0 R
/Type/StructElem/S/mi
>>
endobj
   ...
   ...
112 0 obj
<<
/K [
 ...
120 0 R
 ...
]
/P 110 0 R
/Type/StructElem/S/Formula 
/ID(Math0.1)/T(InlineMath 0.1)
/AF [27 0 R 29 0 R] /A <</O/XML-1.01 >>
>>
endobj
\end{verbatim}
\end{multicols}}%
\removelastskip}% end of box0
\vrule height \ht0 width 0pt 
\par
\removelastskip
\leavevmode%
 (a) PDF coding of the \pdt{/Formula} structure node showing the reference to `Associated Files'
 via the \pdt{/AF} key in object 112. 
 The indirect references (27 and 29) correspond to \pdt{/Filespec} dictionaries,
 as shown in Fig.\,\ref{assoc-files}. 
 (In the coding `\,\dots' indicates parts omitted due to not being relevant to this structure; 
 these portions are discussed in Sect.\,\ref{sect3}.)
 The corresponding \emph{marked content} is specified via the \pdt{/K [ ... ]} numbers (9, 10, 11)
 in the child structure nodes; i.e., objects 114 (\texttt{<mi>}), 116 (\texttt{<mo>}) and 118 (\texttt{<mi>}), 
 which are children of object 121 (\texttt{<mrow>}) under object 120 (\texttt{<math>}).\par
\subfiglabel{1b}\vskip-\baselineskip
\setbox0=\vbox{\hsize=\linewidth
{\scriptsize
\begin{multicols}2
\begin{verbatim}
/mi <</MCID 9
 /ActualText<FEFFD835DC58>/Alt(  k  )
>>BDC
BT
/F30 10.9091 Tf
 [(k)]TJ
ET
EMC
1 0 0 1 6.023 0 cm
/mo <</MCID 10
 /ActualText<FEFF2208>/Alt(  as element of  )
>>BDC
1 0 0 1 3.03 0 cm
BT
/F33 10.9091 Tf
 [(2)]TJ
ET
EMC
1 0 0 1 10.303 0 cm
/mi <</MCID 11
 /ActualText<FEFF211D>/Alt(  real numbers  )
>>BDC
BT
/F42 10.9091 Tf
 [(R)]TJ
ET
EMC
\end{verbatim}
\end{multicols}
}\removelastskip}% end of box0
\vrule height \ht0 width 0pt \par
(b) 
 Portion of the PDF page content stream showing the \pdt{/MCID} numbers (9, 10, 11)
 of the actual content portions of the mathematical expression.
 These correspond to leaf-nodes of the structure tree as presented in part \subfigref[(a)]{1a}.
\vskip-.5\baselineskip
\caption{PDF coding for portions of \subfigref[(a)]{1a} the structure tree 
and \subfigref[(b)]{1b} the page content stream,
corresponding to the mathematics shown as selected in Fig.\,\subfigref{1a}. %\ref{assoc-files}(a).
%The \LaTeX\ source and \MathML\ Supplementary description are included within the \PDF\ document 
%as `Associated Files', as shown in Figure~\ref{assoc-files}.
It is through the \pdt{/MCID} numbers that the association is made to the \pdt{/Formula} structure tag 
for the corresponding piece of mathematical content.
}\label{formula-code}
\end{figure}

\begin{figure}[p]
%   Please do not adjust anything that affects vertical spacing within this Figure.
%   The subfigure captions have been carefully positioned over the large image.
\subfiglabel{2a}\vspace{-12pt}%
\subfigure[Listing of attachments, indicating how one is associated to some inline content.]{%
%\centering
 \centerline{%
  \includegraphics[width=.75\textwidth,viewport=0 0 1450 730,clip]{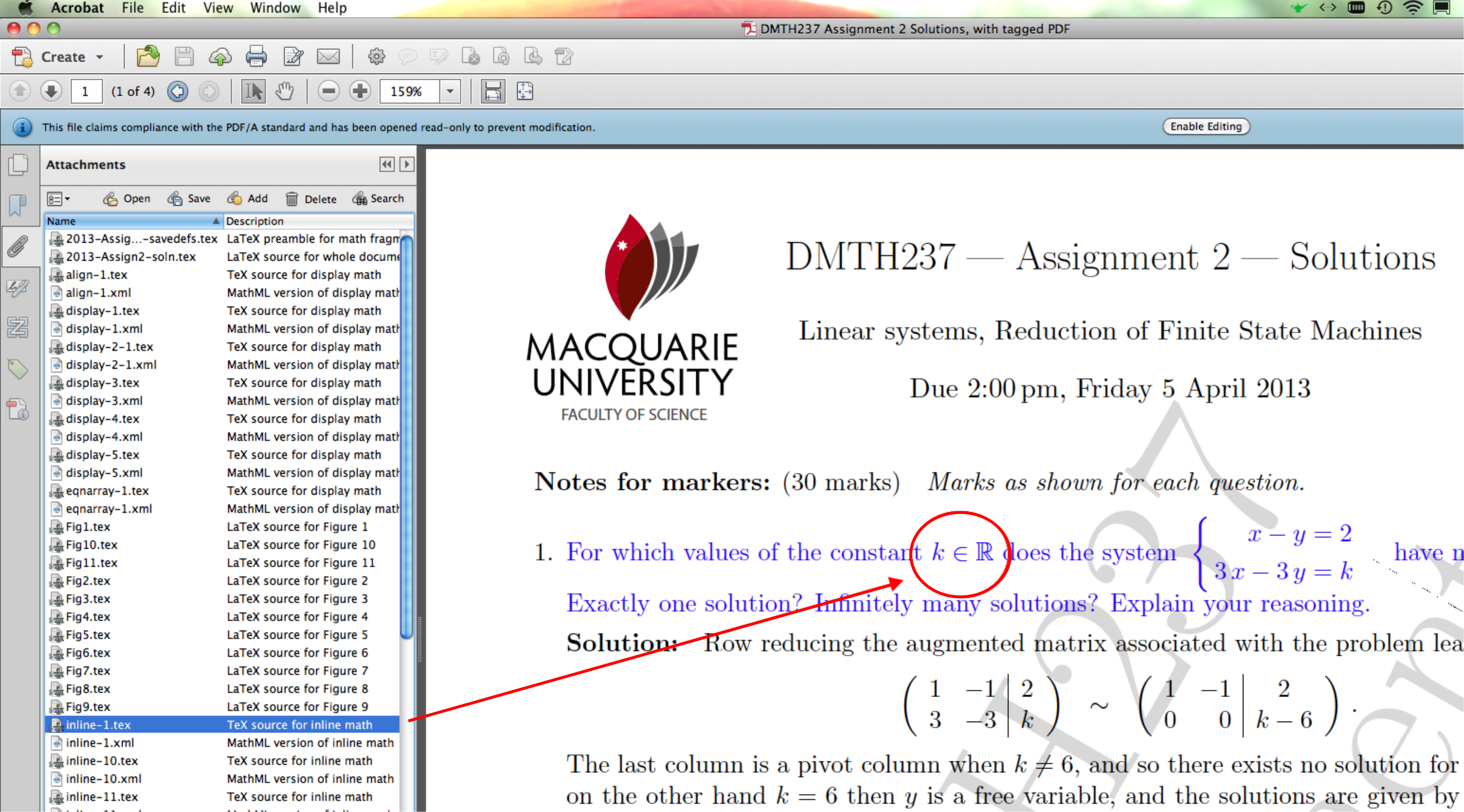}}%
}\par
\removelastskip %\par
\subfiglabel{2b}%
\vskip-\baselineskip
\setbox0=\hbox{\includegraphics[]{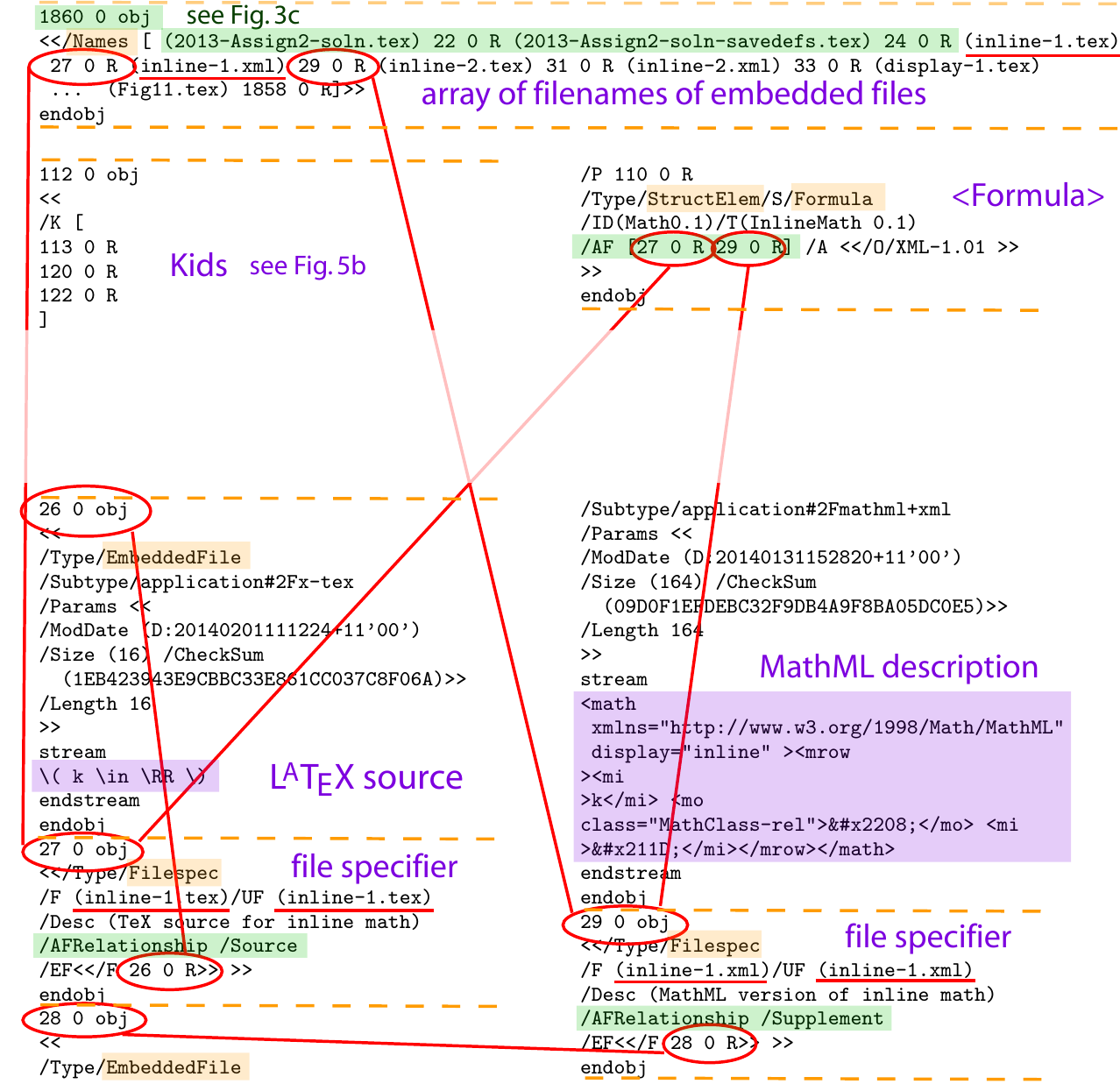}}%-opt-2}}% don't use the 'optimized' version; some stream errors
\setbox0=\hbox{\kern-15pt\lower\ht0\box0}%
\ht0=0pt \dp0=0pt \box0\par
\setbox0=\vbox{\hsize=\linewidth
{\scriptsize
\begin{verbatim}
1860 0 obj
<</Names [ (2013-Assign2-soln.tex) 22 0 R (2013-Assign2-soln-savedefs.tex) 24 0 R (inline-1.tex) 
 27 0 R (inline-1.xml) 29 0 R (inline-2.tex) 31 0 R (inline-2.xml) 33 0 R (display-1.tex) 
 ...  (Fig11.tex) 1858 0 R]>>
endobj
\end{verbatim}
}%
%\removelastskip\smallskip
{\scriptsize
\begin{multicols}2
\begin{verbatim}
112 0 obj
<<
/K [
113 0 R
120 0 R
122 0 R
]
/P 110 0 R
/Type/StructElem/S/Formula 
/ID(Math0.1)/T(InlineMath 0.1)
/AF [27 0 R 29 0 R] /A <</O/XML-1.01 >>
>>
endobj
\end{verbatim}
\end{multicols}}%
\removelastskip
}% end of box0
\vrule height \ht0 width 0pt \par
\removelastskip
\vskip-.5\baselineskip
\leavevmode
\quad (b) 
A portion of the PDF coding of the \pdt{/Names} array (upper) for embedded files, 
and coding of the \pdt{/Formula} structure element (lower), 
showing the \pdt{/AF} key with array of two `Associated' files given as indirect references.
The \pdt{/Names} array is used to produce the listing in \subfigref[(a)]{2a} 
and provides indirect links to \pdt{/Filespec} entries,
as shown in \subfigref[(c)]{2c}.\par
%\removelastskip\smallskip
\subfiglabel{2c}%
\vskip-\baselineskip
\setbox0=\vbox{\hsize=\linewidth
{\scriptsize
\begin{multicols}2
\begin{verbatim}
26 0 obj
<<
/Type/EmbeddedFile 
/Subtype/application#2Fx-tex
/Params <<
/ModDate (D:20140201111224+11'00') 
/Size (16) /CheckSum
  (1EB423943E9CBBC33E861CC037C8F06A)>> 
/Length 16        
>>
stream
\( k \in \RR \) 
endstream
endobj
27 0 obj
<</Type/Filespec 
/F (inline-1.tex)/UF (inline-1.tex)
/Desc (TeX source for inline math)
/AFRelationship /Source 
/EF<</F 26 0 R>> >>
endobj
28 0 obj
<<
/Type/EmbeddedFile 
/Subtype/application#2Fmathml+xml
/Params <<
/ModDate (D:20140131152820+11'00') 
/Size (164) /CheckSum
  (09D0F1EFDEBC32F9DB4A9F8BA05DC0E5)>> 
/Length 164       
>>
stream
<math 
 xmlns="http://www.w3.org/1998/Math/MathML" 
 display="inline" ><mrow
><mi 
>k</mi> <mo 
class="MathClass-rel">&#x2208;</mo> <mi 
>&#x211D;</mi></mrow></math>
endstream
endobj
29 0 obj
<</Type/Filespec 
/F (inline-1.xml)/UF (inline-1.xml)
/Desc (MathML version of inline math)
/AFRelationship /Supplement 
/EF<</F 28 0 R>> >>
endobj
\end{verbatim}
\end{multicols}}%
\removelastskip
}% end of box0
\vrule height \ht0 width 0pt \par
%\removelastskip\smallskip
\leavevmode
\quad (c) 
PDF source of \pdt{/Filespec} dictionaries and \pdt{/EmbeddedFile} objects 
which hold the streams of \LaTeX\ and \MathML\ coding for the mathematical content 
indicated in \subfigref[(a)]{2a}.
\par
\caption{Embedded files associated with a \pdt{/Formula} structure element.}
\label{assoc-files}
\end{figure}

\begin{figure}[hpbt]
%   Please do not adjust anything that affects vertical spacing within this Figure.
%   The subfigure captions have been carefully positioned over the large image.
\subfiglabel{3a}%
\vspace{-18pt}%
\setbox0=\hbox{\includegraphics[]{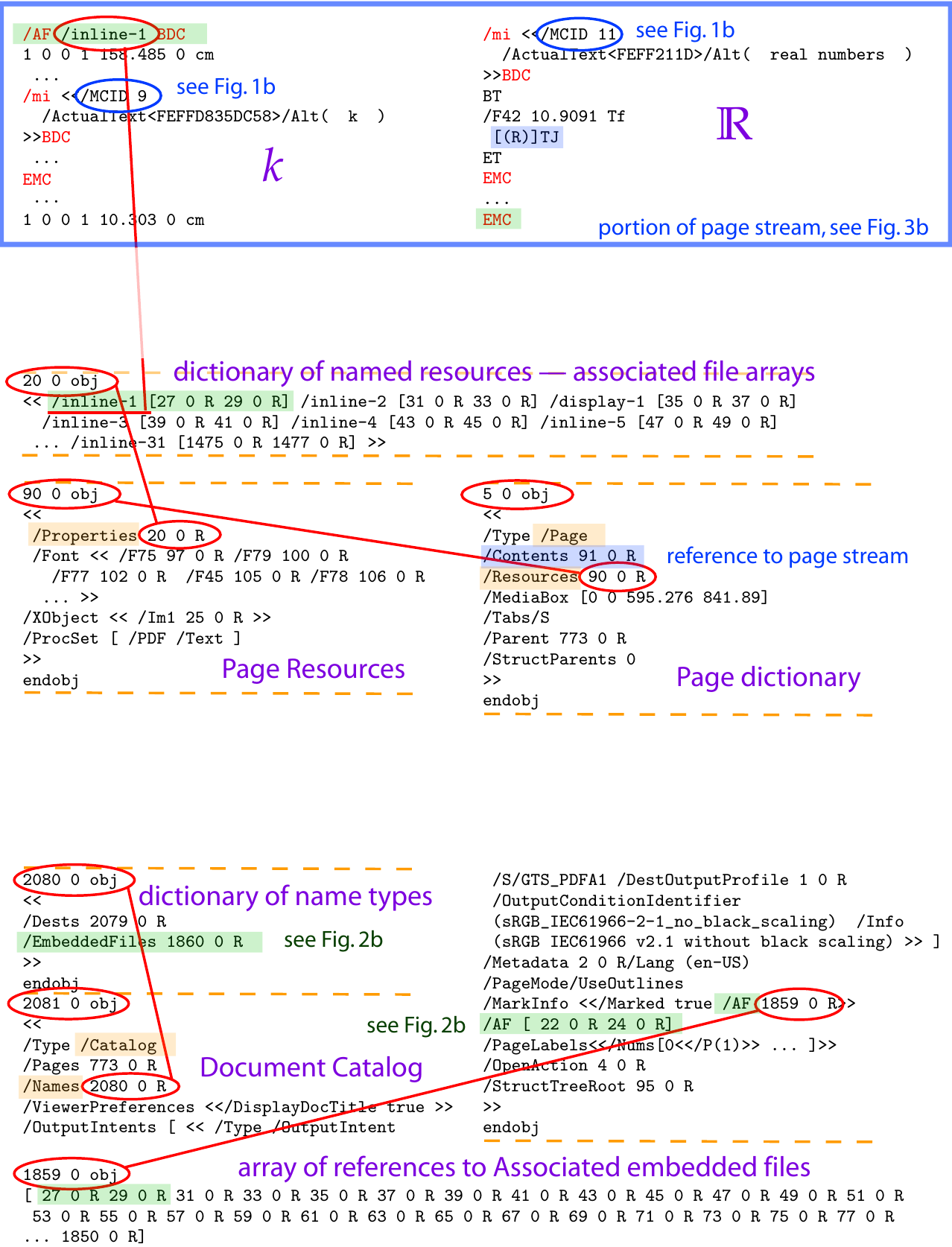}}%
\setbox0=\hbox{\kern-10pt\lower\ht0\box0}%
\ht0=0pt \dp0=0pt \box0\par
\setbox0=\vbox{\hsize=\linewidth
{\scriptsize
\begin{multicols}2
\begin{verbatim}
/AF /inline-1 BDC
1 0 0 1 158.485 0 cm
 ...
/mi <</MCID
  9 /ActualText<FEFFD835DC58>/Alt(  k  )
>>BDC
 ...
 ...
EMC
1 0 0 1 10.303 0 cm
/mi <</MCID 11
  /ActualText<FEFF211D>/Alt(  real numbers  )
>>BDC
BT
/F42 10.9091 Tf
 [(R)]TJ
ET
EMC
...
EMC
\end{verbatim}
\end{multicols}}%
%\removelastskip
}% end of box0
\vrule height \ht0 width 0pt \par
\removelastskip
%\vskip-.5\baselineskip
\leavevmode
\quad (a) 
A portion of the PDF content stream associating content
with embedded files for the mathematical expression indicated in Fig.\,\subfigref{2a}, %\ref{assoc-files}(a),
using a \emph{marked content} tag \pdt{/AF} to refer to a \emph{named resource} \pdt{/inline-1}.
All content down to the final \texttt{EMC} is associated.\par
\subfiglabel{3b}\vskip-\baselineskip
\setbox0=\vbox{\hsize=\linewidth
{\scriptsize
\begin{verbatim}
20 0 obj
<< /inline-1 [27 0 R 29 0 R] /inline-2 [31 0 R 33 0 R] /display-1 [35 0 R 37 0 R] 
  /inline-3 [39 0 R 41 0 R] /inline-4 [43 0 R 45 0 R] /inline-5 [47 0 R 49 0 R] 
 ... /inline-31 [1475 0 R 1477 0 R] >>
\end{verbatim}
\begin{multicols}2%
\begin{verbatim}
90 0 obj
<<
 /Properties 20 0 R
 /Font << /F75 97 0 R /F79 100 0 R
   /F77 102 0 R  /F45 105 0 R /F78 106 0 R 
  ... >>
/XObject << /Im1 25 0 R >>
/ProcSet [ /PDF /Text ]
>>
endobj
\end{verbatim}%
\begin{verbatim}
5 0 obj
<<
/Type /Page
/Contents 91 0 R
/Resources 90 0 R
/MediaBox [0 0 595.276 841.89]
/Tabs/S
/Parent 773 0 R
/StructParents 0
>>
endobj
\end{verbatim}%
\end{multicols}}%
%\removelastskip
}% end of box0
\vrule height \ht0 width 0pt \par
%\removelastskip\smallskip
\leavevmode
\quad (b) 
A portion of the \pdt{/Properties} dictionary (upper, object 20)
which is linked to a \pdt{/Page} object (lower right, object 5) 
via its \pdt{/Resources} key (see lower left, object 90).
Thus a name (such as \pdt{/inline-1}) is associated with an array of \pdt{/Filespec} references
(viz. \verb|[27 0 R 29 0 R]|), which lead to the \LaTeX\ and \MathML\ files seen
in Fig.\,\subfigref{2c}.%\ref{assoc-files}(c). 
\par
\subfiglabel{3c}\vskip-\baselineskip
\setbox0=\vbox{\hsize=\linewidth
{\scriptsize
\begin{multicols}2%
\begin{verbatim}
2080 0 obj
<<
/Dests 2079 0 R
/EmbeddedFiles 1860 0 R
>>
endobj
2081 0 obj
<<
/Type /Catalog
/Pages 773 0 R
/Names 2080 0 R
/ViewerPreferences <</DisplayDocTitle true >> 
/OutputIntents [ << /Type /OutputIntent 
 /S/GTS_PDFA1 
 /DestOutputProfile 1 0 R /OutputConditionIdentifier 
 (sRGB_IEC61966-2-1_no_black_scaling)  /Info
 (sRGB IEC61966 v2.1 without black scaling) >> ]
/Metadata 2 0 R/Lang (en-US)
/PageMode/UseOutlines
/MarkInfo <</Marked true /AF 1859 0 R>>
/AF [ 22 0 R 24 0 R]
/PageLabels<</Nums[0<</P(1)>> ... ]>>
/OpenAction 4 0 R
/StructTreeRoot 95 0 R
>>
endobj
\end{verbatim}%
\end{multicols}
\begin{verbatim}
1859 0 obj
[ 27 0 R 29 0 R 31 0 R 33 0 R 35 0 R 37 0 R 39 0 R 41 0 R 43 0 R 45 0 R 47 0 R 49 0 R 51 0 R 
 53 0 R 55 0 R 57 0 R 59 0 R 61 0 R 63 0 R 65 0 R 67 0 R 69 0 R 71 0 R 73 0 R 75 0 R 77 0 R 
... 1850 0 R]
\end{verbatim}%
}%
\removelastskip
}% end of box0
\vrule height \ht0 width 0pt \par
%\vspace{-.7\baselineskip}
\leavevmode
\quad (c) 
The document's \pdt{/Catalog} (object 2081) indicates presence of embedded files via the \pdt{/Names} key (object 2080). 
This references the array (object 1860 in Fig.\,\subfigref{2b}), %\ref{assoc-files}b), 
to establish the correspondence between filenames and \pdt{/Filespec} dictionaries.
Embedded files which are `Associated' to content portions are also listed in an array (object 1859) 
referenced from the \pdt{/AF} key in the \pdt{/MarkInfo} dictionary.\par
\vspace{-.7\baselineskip}
\caption{Embedded files associated with specific content.}\label{assoc-cont}%
\end{figure}

The overall structure of an uncompressed PDF file consists of:
\begin{enumerate}[(a)]
\item
 a collection of numbered \emph{\bfseries objects}: written as  \texttt{<}\textit{num}\texttt{> 0 obj }\dots\ \texttt{endobj}\ 
 where the `\,\dots' can represent many, many lines of textual (or binary) data starting on a new line after \texttt{obj}
 and with \texttt{endobj} on a line by itself.
 The numbering need not be sequential and objects may appear in any order.
 An \emph{\bfseries indirect reference} sequence of the form  \texttt{<}\textit{num}\texttt{> 0 R}\footnote{%
  The `\texttt{0}' is actually a \textit{revision number}.
  In a newly constructed PDF this will always be~0;
  but with PDF editing software, higher numbers can result from edits.}
  is used where data from one object is required when processing another.
 A \emph{cross-reference table} (described next), allows an object and its data to be located precisely.
 Such indirect references are evident throughout the coding portions of Figures\,\ref{formula-code}--\ref{access-tags}.
 \item
  the \emph{\bfseries cross-reference table}:
  listing of byte-offsets to where each numbered object occurs within the uncompressed PDF file,
  together with a linked listing of unused object numbers.
  (Unused numbers are available for use by PDF editing software.)
 \item
  the \emph{trailer}, including: 
 \begin{inparaenum}[(i)]
 \item
  total number of objects used; 
 \item
  reference to the document's \pdt{/Catalog}, %or root node, 
  see Fig.\,\subfigref{3c};
 \item
  reference to the \pdt{/Info} dictionary, containing file properties  (i.e., basic metadata);
 \item
  byte-offset to the cross-reference table;
 \item
  encryption and decryption keys for handling compression;
 \item
  end-of-file marker.
 \end{inparaenum} 
 \end{enumerate}
Thus the data in a PDF file is contained within the collection of objects, 
using the cross-reference table to precisely locate those objects. 
A PDF browser uses the \pdt{/Catalog} object (e.g., object 2081 in Fig.\,\subfigref{3c}) %\ref{assoc-cont}c)
to find the list of \pdt{/Page} objects (e.g., object 5 in Fig.\,\subfigref{3b}), %\ref{assoc-cont}b), 
each of which references a \pdt{/Contents} object. 
This provides each page's \emph{\bfseries contents stream} of graphics commands,
which give the details of how to build the visual view of the content to be displayed.
A small portion of the page stream for a particular page is shown 
in Figures~\subfigref{1b}, %\ref{formula-code}b, 
\subfigref{3a}, %\ref{assoc-cont}a, 
\subfigref{5a}. %\ref{access-tags}a.

\emph{\bfseries Character strings} are used in PDF files in various ways; 
most commonly for ASCII strings, in the form \texttt{(}\,\dots\texttt{)}; see
Figures~\subfigref{1a}, \subfigref{1b}, %\ref{formula-code}a,b 
\subfigref{2b}, \subfigref{2c}, %\ref{assoc-files}b,c 
\subfigref{3a}, \subfigref{3c}, %\ref{assoc-cont}a,c 
and \subfigref{5a}. %\ref{access-tags}a.
Alternatively, a hexadecimal representation with byte-order mark \texttt{<FEFF}\dots\texttt{>} can be used, 
as in Figures~\subfigref{1b}, %\ref{formula-code}b, 
\subfigref{3a}, %\ref{assoc-cont}a, 
\subfigref{5a}. %\ref{access-tags}a.
This is required particularly for Unicode characters above position 255, 
with `surrogate pairs' used for characters outside the basic plane, as with the $k$ variable name in those figures. %and other kinds of encoded data.
Below 255 there is also the possibility of using 3-byte octal codes within the \texttt{(}\,\dots\texttt{)} string format; 
see \Octalfootmark\ in Sect.\,\ref{sect3}.
For full details, see \S7.3.4 of PDF Specifications \cite{PDF17,ISO32000}.

\emph{\bfseries PDF names} of the form \pdt{/}$\BNF{name}$, usually using ordinary letters,
have a variety of uses, including
%These are used for: 
\begin{inparaenum}[(i)]
\item
 \emph{\bfseries tag-names} in the content stream (Figures~\subfigref{1b}, %\ref{formula-code}b, 
 \subfigref{3a}, %\ref{assoc-cont}a, 
 \subfigref{5a}); %\ref{access-tags}a);
\item
 identifiers for \emph{\bfseries named resources} (Fig.\,\subfigref{3b} %\ref{assoc-cont}b 
 within object 20 and in the \pdt{/AF} tagging shown in Fig.\,\subfigref{3a}); %\ref{assoc-cont}a); 
and extensively as
\item
dictionary \emph{\bfseries keys} (in all the Figures~%1, 2, 3, 5) 
\ref{formula-code}, \ref{assoc-files}, \ref{assoc-cont}, \ref{access-tags}) 
and frequently as dictionary \emph{\bfseries values} (see below). 
\end{inparaenum}

\medskip\noindent
Other common structures used within PDF objects are as follows.
\begin{enumerate}[(i)]
\item
 \emph{\bfseries arrays}, represented as 
 \texttt{[}$\BNF{item}\ \BNF{item}\ \ldots\ \BNF{item}$\texttt{]},
 usually with similar kinds of $\BNF{item}$, (see e.g., Figures~\subfigref{1a}, %\ref{formula-code}a, 
 \subfigref{3b}, %\ref{assoc-cont}b, 
 \subfigref{3c}) %\ref{assoc-cont}c)
 or alternating kinds (e.g., the filenames array of Fig.\,\subfigref{2b}. %\ref{assoc-files}b).
\item
 \emph{\bfseries dictionaries} 
  of \emph{key--value pairs}, similar to alternating arrays, but represented as 
 \texttt{<<}$\BNF[_1]{key}\ \BNF[_1]{value}\ \BNF[_2]{key}\ \BNF[_2]{value}$\ \dots\ \texttt{>>}.
 The $\BNF{key}$ is always a \emph{PDF\,name} whereas the $\BNF{value}$ may be any other element
 (e.g., string, number, name, array, dictionary, indirect reference).
 The \emph{key--value pairs} may occur in any order, with the proviso that if the same $\BNF{key}$ occurs more than
 once, it is the first instance whose $\BNF{value}$ is used.
 A \pdt{/Type} key, having a \emph{PDF\,name} as value, is not always mandatory; 
 but when given, one refers to the dictionary object as being of the type of this name.
 See Figures~\subfigref{1a}, %\ref{formula-code}a, 
 \subfigref{2b}, \subfigref{2c}, %\ref{assoc-files}b,c 
 \subfigref{3b}, \subfigref{3c} %\ref{assoc-cont}b,c 
 and \subfigref{5b} %\ref{access-tags}b 
 for examples.
\item
 \emph{\bfseries stream objects} 
 consist of a dictionary followed by an arbitrarily-long delimited \emph{stream} of data,
 having the form  \texttt{<<\,\dots\,>> stream \dots\ endstream}, with the \texttt{stream} and \texttt{endstream}
 keywords each being on a separate line by themselves (see objects 26 and 28 in Fig.\,\subfigref{2c}). %\ref{assoc-files}c).
 The dictionary must include a \pdt{/Length} key, whose value is the integer number of bytes within the data-stream.
 With the length of the data known, between the keywords on separate lines, there is no need for any escaping
 or special encoding of any characters, as is frequently needed in other circumstances and file-formats.
 See \S7.3.8 of \cite{PDF17,ISO32000} for more details;
 e.g., how compression can be used.
\item
 \emph{\bfseries graphics operators} which place font characters into the visual view %printed page 
 occur inside a page contents stream, 
 within portions delimited by \texttt{BT} \dots\ \texttt{ET} (abbreviations for Begin/End\,Text); 
 see Figures~\subfigref{1b}, %\ref{formula-code}b, 
 \subfigref{3a}, %\ref{assoc-cont}c, 
 \subfigref{5a}. %\ref{access-tags}a.
 These include coding \pdt{/}$\BNF{fontname}\ \BNF{size}$\ \texttt{Tf } for selecting the (subsetted) font, 
 scaled to a particular size, and \texttt{[}\,$\BNF{string}$\,\texttt{]TJ } 
 for setting the characters of the string with the previously selected font. 
 See \S9.4 of \cite{PDF17,ISO32000} for a complete description of the available text-showing
 and text-positioning operators.
\end{enumerate}
Dictionaries and arrays can be nested; that is, the $\BNF{value}$ of a dictionary item's $\BNF{key}$ may well
be another dictionary or array, as seen in objects 20 and 90 within Fig.\,\subfigref{3b}. %\ref{assoc-cont}b.
Similarly one or more $\BNF{item}$s in an array could well be a dictionary, another array, 
or an indirect reference (regarded as a `pointer' to another object).

With the use of \emph{PDF\,names}, \emph{objects}, and \emph{indirect references}
a PDF file is like a self-contained web of interlinked information,
with names chosen to indicate the kind of information referenced or how that information should be used. 

The use of objects, dictionaries (with key--value pairs) 
and indirect references makes for a very versatile container-like file format. 
If PDF reader software does not recognise a particular key occurring within a particular type
of dictionary, then both the key and its value are ignored.
When that value is an indirect reference to another object, such as a \emph{stream object},
then the data of that stream may never be processed, so does not contribute to the view being built.
Thus PDF producing or editing software may add whatever objects it likes, for its own purposes,
without affecting the views that other PDF reading software wish to construct.
This should be contrasted with HTML and XML when a browser does not recognise a custom tag.
There that tag is ignored, together with its attributes, 
but any content of that tag \emph{must still be handled}.

It is this feature of the PDF language which allows different reader software to  support different
features, and need not use all of the information contained within a PDF file.
For example, some browsers support attachments; others do not.
A PDF format specification now consists mostly of saying which tags and dictionary keys
\emph{must} be present, what others are allowed, 
and how the information attached to these keys and tags is intended to be used.
Hence the proliferation of different standards: PDF/A, PDF/E, PDF/VT, PDF/UA, PDF/X, 
perhaps with several versions or revisions,
intended for conveying different kinds of specialised information most relevant within specific contexts.
%A document can be compatible with multiple standards. 
%though just one is declared for verification. 

\subsection{Tagging within PDF documents}\label{PDFtagging}%
\noindent
Two types of tagging can be employed within PDF files.
`Tagged PDF' documents use both,
with content tags connected as leaf-nodes of the structure tree.
\paragraph*{Tagging of content} is done as
%within the page contents stream, as 
\pdt{/}$\BNF{tag}\ \BNF{dict}$\texttt{ BDC }\dots\ \texttt{EMC}
within a contents stream.
Here the \texttt{BDC} and \texttt{EMC} stand for `Begin Dictionary Content' 
and `End Marked Content' respectively, 
with the $\BNF{dict}$ providing key-value pairs that specify `properties' of the \emph{\bfseries marked content},
much like `attributes' in XML or HTML tagging\footnote{Henceforth we use the term `attribute', rather than `property'.}.
The $\BNF{tag}$ can in principle be any \emph{PDF\,name}; 
however, in \S14.6.1 of the specifications \cite{PDF17,ISO32000} it stipulates that
``All such tags shall be registered with Adobe Systems (see Annex~E) 
to avoid conflicts between different applications marking the same content stream.''
Thus one normally uses a standard tag, such as \pdt{/Span}, 
or in the presence of structure tagging (see below) choose the same tag name as for the parent structure node.
Figures~\subfigref{1b}, %\ref{formula-code}b, 
\subfigref{3a}, %\ref{assoc-cont}a, 
\subfigref{5a} %\ref{access-tags}a 
show the use of Presentation-MathML content tag names,
which are expected to be supported in PDF\,2.0 \cite{PDF20}.
Typical attributes are the \pdt{/ActualText} and \pdt{/Alt} strings, which allow replacement text to be used
when content is extracted from the document using \textsf{Copy}/\textsf{Paste} or as `Accessible Text' respectively.
The \pdt{/MCID} attribute allows \emph{marked content} to be linked to document structure, as discussed below.
A variant of this %kind of 
tagging uses a \emph{named resource} for the $\BNF{dict}$ element. 
This is illustrated with \pdt{/AF} content tagging in Sect.\,\ref{embed-cont}.

\paragraph*{Tagging of structure} 
requires building a tree-like structural description of a document's contents,
in terms of Parts, Sections, Sub-sections, Paragraphs, etc. 
and specialised structures such as Figures, Tables, Lists, List-items, and more \cite[\S14.8.4]{PDF17,ISO32000}.
Each \emph{\bfseries structure node} is a dictionary of type \pdt{/StructElem} having keys \pdt{/S} for the structure type,
\pdt{/K} an array of links to any child nodes (or Kids) including \emph{marked content} items,
and \pdt{/P} an indirect reference to the parent node.
Optionally there can be a \pdt{/Pg} key specifying an indirect reference to a \pdt{/Page} dictionary,
when this cannot be deduced from the parent or higher ancestor. %  --- structure is not confined within page boundaries.
Also, the \pdt{/A} key can be used to specify attributes for the structure tag when the document's contents are exported in various formats; 
e.g., using  `\textsf{Save As Other ... XML 1.0}' export from Adobe's `Acrobat Pro' browser/editor.
Fig.\,\subfigref{1a} %\ref{formula-code}a 
shows the MathML tagging of some inline mathematical content.
The tree structure is indicated with lines connecting nodes to their kids; 
reverse links to parents are not drawn, as this would unduly clutter the diagram.
Other keys, such as \pdt{/ID} and \pdt{/T} can provide an identifier and title, 
for use primarily in editing software to locate specific nodes within appropriately ordered listings. 

\medskip
The link between structure and \emph{marked content} (as leaf-nodes to the structure tree, say) 
is established using the \pdt{/MCID} number attribute.
A numeric integer entry in the \pdt{/K} Kids array corresponds to an \pdt{/MCID} number occurring
within the contents stream for that page specified via a \pdt{/Pg} entry, 
either of the structure node itself or the closest of its ancestors having such a key.
Fig.\,\subfigref{1b} %\ref{formula-code}b 
shows this linking via \pdt{/MCID} with lines drawn to the corresponding
structure nodes shown in Fig.\,\subfigref{1a}. %\ref{formula-code}a.
The interplay of structure with content was addressed in the author's paper \cite{DML2009},
with Figure 1 of that paper giving a schematic view of the required PDF structural objects.

\section{`Associated Files', carrying \LaTeX\ and \MathML\ views of mathematical content.}%
\label{sect2}%
There are several ways in which file attachments may be associated with specific portions of a PDF document, 
using the `Associated Files' technique \cite[Annex~E]{pdfA3}. 
The file is embedded/attached and then associated, by a method, either to:
\begin{enumerate}[\hss(i)\hss]
\item
  the document as a whole %\cite[\S14.13.2]{PDF20}
 \cite[\S E.3]{pdfA3}, \cite[\S14.13.2]{PDF20}
 --- e.g. the full \LaTeX\ source, or preamble file used when converting snippets
  of mathematical content into a \MathML\ presentation of the same content;
\item
  a specific page within the document %\cite[\S14.13.3]{PDF20}
 \cite[\S E.4]{pdfA3}, \cite[\S14.13.3]{PDF20}
 or to a (perhaps larger) logical document part using PDF\,2.0 \cite[\S14.13.7]{PDF20};
\item
 graphic objects in a content stream %\cite[\S14.13.4]{PDF20}
 \cite[\S E.5]{pdfA3}, \cite[\S14.13.4]{PDF20}
  --- when structure is available, this is not the preferred method\footnote{%
In the PDF/A-3 specifications %\cite[\S14.13.4]{PDF20}
 \cite[\S E.5]{pdfA3} %, \cite[\S14.13.4]{PDF20} 
 the final paragraph %in this sub-section 
 explicitly states 
``When writing a PDF,  the use of structure (and thus associating the \pdt{/AF} with the structure element,
 see \cite[\S E.7]{pdfA3}) is preferred instead of the use of explicit marked content.''
  with a corresponding statement also in \cite[\S14.13.4]{PDF20}.};
\item
 a structure node %\cite[\S14.13.5]{PDF20}
 \cite[\S E.7]{pdfA3}, \cite[\S14.13.5]{PDF20} 
 such as \pdt{/Figure}, \pdt{/Formula}, \pdt{/Div}, %\pdt{/Para}, 
 etc.
\item
 an \pdt{/XObject} %\cite[\S14.13.6]{PDF20}
 \cite[\S E.6]{pdfA3}, \cite[\S14.13.6]{PDF20}
 such as an included image of a formula or other mathematical/technical/diagrammatic content;
\item
 an annotation %\cite[\S14.13.8]{PDF20}
 \cite[\S E.8]{pdfA3} --- but this method can be problematic with regard to validation 
 for PDF/A \cite[\S6.3]{pdfA3}, and PDF/UA \cite[\S7.18]{PDF-UA1} standards\footnote{%
The method of indicating an attachment with a `thumb tack' annotation located at a specific point within a document, 
is deprecated in the PDF/A-3 standard, as it does not provide a proper method to associate with the portion of content.
Besides, the appearance of such thumb-tacks all over paragraphs containing inline mathematics is, well, \dots\ \ \ 
\dots downright ugly.}.
\end{enumerate}
Fig.\,\subfigref{2a} %\ref{assoc-files}a 
shows how attachments are presented within a separate panel of 
a browser window, using information from an array of filenames; see Fig.\,\subfigref{2b}. %\ref{assoc-files}b.
This is independent of the page being displayed, so the array must be referenced from the document level.
This is seen in Fig.\,\subfigref{3c} %\ref{assoc-cont}c 
using the \pdt{/Names} key of the \pdt{/Catalog} dictionary,
which references object 2080, 
whose \pdt{/EmbeddedFiles} key then references the filenames array (object 1860 in Fig.\,\subfigref{2b}). %\ref{assoc-files}b).
One can also see in Fig.\,\subfigref{2b} %\ref{assoc-files}b 
how each filename precedes an indirect reference 
to the \pdt{/Filespec} dictionary \cite[\S7.11.3, Tables 44 and 45]{PDF20} for the named file; 
see Fig.\,\subfigref{2c}. %\ref{assoc-files}c.
This dictionary contains a short description (\pdt{/Desc}) of the type of content as well as the filename to use
on disk, and a link via the \pdt{/EF} key to the actual \pdt{EmbeddedFile} stream object.

That a file is `Associated' is indicated by the \pdt{/AFRelationship} key, 
whose value is a \emph{PDF name} indicating how the file is related to visible content.
Options here are \pdt{/Source} as used with the \LaTeX\ source coding, 
or \pdt{/Supplement} as used with the \MathML\ description.
Other possibilities are \pdt{/Data} (e.g., for tabular data) 
and \pdt{/Alternative} for other representations such as audio, a movie, projection slides
or anything else that may provide an alternative representation of the same content.
\pdt{/Unspecified} is also available as a non-specific catch-all.

Not all attachments need be `Associated' and conversely not all `Associated Files' need be displayed
in the `Attachments' panel, so there is another array (object 1859) as shown in Fig.\,\subfigref{3c}, %\ref{assoc-cont}c,
linked to the \pdt{/MarkInfo} sub-dictionary of the \pdt{/Catalog} dictionary.
Files associated with the document as a whole, as in method~(i) above,
link via the \pdt{/AF} key in the \pdt{/Catalog} dictionary (see Fig.\,\subfigref{3c}).%\ref{assoc-cont}c).

For the \LaTeX\ source of a mathematical expression method (iv) is preferred, 
provided structure tagging is present within the PDF.
This is discussed below in Sect.\,\ref{embed-struct}.
Method (iii) also works, provided the expression is built from content confined to a single %physical 
page. %, 
%whether inline or displayed.
This is described in Sect.\,\ref{embed-cont}.

As `Associated Files' have only been part of published PDF/A standards~\cite{pdfA3} %\cite{ISO32000},
since late 2012, %and the specifications for PDF\,2.0 \cite{PDF20} is yet to be finalised, 
it may be some time before PDF readers provide a good interface for `Associated Files',
beyond using the `\textsf{Attachments}' pane. % --- provided the reader even supports attachments\footnote{%
%Apple Inc.'s software typically does not!}.
This ought to include interfaces to view the contents of attached files, do searching within the files,
and make the file's contents available to assistive technology.
One possible way to display this association is apparent in earlier work~\cite{TUG2002},
whereby a bounding rectangle appears as the mouse enters the appropriate region.

\subsection{Embedded files associated with structure}\label{embed-struct}%
\noindent
With an understanding of how structure tagging works, as in Sect.\,\ref{PDFtagging},
then associating files to structure is simply a matter of including an \pdt{/AF} key in the structure node's
dictionary, as shown in Figures~\subfigref{1a} %\ref{formula-code}a 
and~\subfigref{2b}. %\ref{assoc-files}b.
The value for this key is an array of indirect references to \pdt{/Filespec} objects for the relevant files.

There is nothing in the content stream in Fig.\,\subfigref{1b} %\ref{formula-code}b 
to indicate that there is a file associated with this structure node.
Rather the browser, knowing that `Associated' files are present, 
needs to have gone through some pre-processing to first locate the node (if any) to which it is associated,
then trace down the structure tree to the deepest child nodes  (objects 114, 116, 118).
From their \pdt{/K} entries (viz., 9, 10, 11 resp.),
the relevant \emph{marked content} in the page's contents stream is located using these \pdt{/MCID} numbers.

\subsection{Embedded files associated with content}\label{embed-cont}%
With an understanding of how content tagging works, as in Sect.\,\ref{PDFtagging},
and the fact that \emph{marked content} operators may be nested,
then associating files to content is also quite simple.
One simply uses an \pdt{/AF} tag within the page's content stream
with \texttt{BDC} \dots\ \texttt{EMC} surrounding the content to be marked,
as shown in Fig.\,\subfigref{3a}. %\ref{assoc-cont}a.
This employs the \emph{named resource} variant (here \pdt{/inline-1})
to indicate the array of `Associated' files.
Fig.\,\subfigref{3b} %\ref{assoc-cont}b 
shows how this name is used as a key (in dictionary object 20) 
having as value an array of indirect references to \pdt{/Filespec} objects (27 and 29).
These resources can be specific to a particular page dictionary (object 5),
but in the example document\Exfootmark\ the \emph{named resources} are actually made available to all pages, 
since this accords with not including multiple copies of files when a mathematical expression is used repeatedly.

Finally Fig.\,\subfigref{3c} %\ref{assoc-cont}c 
shows the coding required when embedded files, 
some of which may also be associated to content or structure, are present within a PDF document.
One sees that the array (object 1859) of indirect object references in the lower part of Fig.\,\subfigref{3c} %\ref{assoc-cont}c 
refer to the same \pdt{/Filespec} objects  (27 and 29) 
as the \emph{named resources} (object 20) in the upper part of Fig.\,\subfigref{3b}. %\ref{assoc-cont}b.
These are the same references using \pdt{/AF} keys seen in Fig.\,\subfigref{1b} %\ref{formula-code}b 
and Fig.\,\subfigref{2b} %\ref{assoc-files}b
to the objects themselves in Fig.\,\subfigref{2c}.%\ref{assoc-files}c.

This mechanism makes it easier for a PDF reader to determine that there are files associated to
a particular piece of content, by simply encountering the \pdt{/AF} tag linked with a \emph{named resource}.
This should work perfectly well with a PDF file that is not fully tagged for structure.
However, if the content is extended (e.g., crosses a page-boundary) then it may be harder for
a PDF writer to construct the correct content stream, properly tagging two or more portions.

\begin{figure}[p]
\subfiglabel{4a}%\vskip-\baselineskip
\centering
\subfigure[Selection across mathematical content]{\includegraphics[width=.825\textwidth]{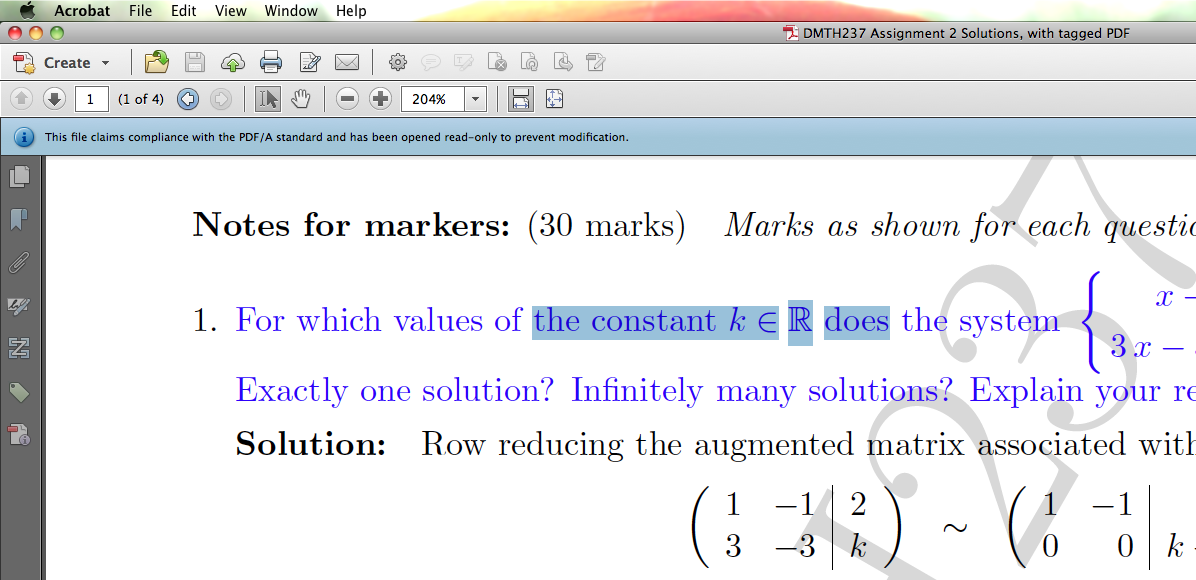}}\par
\bigskip
\subfiglabel{4b}\vskip-\baselineskip
\subfigure[Pasted text from the selection]{\includegraphics[width=.825\textwidth]{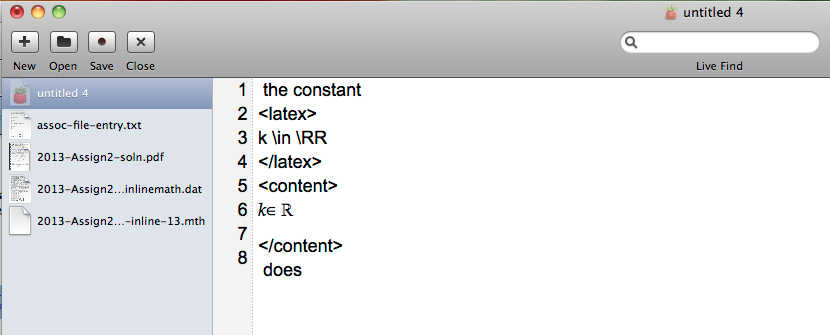}}\par
\bigskip
\subfiglabel{4c}\vskip-\baselineskip
%\subfigure[Tag structure by selection in the `Tags' tree.]{\includegraphics[width=.75\textwidth]{tag-selection}}\par
\subfigure[Access-tags selected in the `Tags' tree, to show `fake spaces'.]{\includegraphics[width=.825\textwidth]{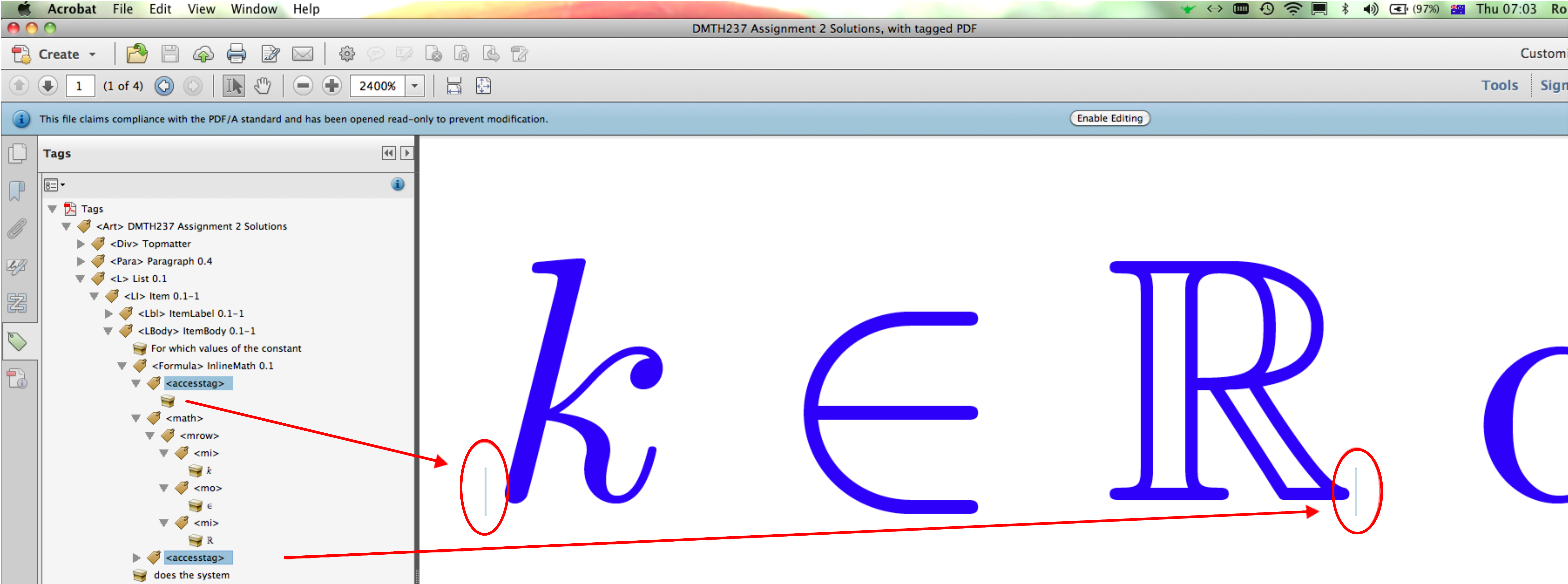}}\par
\caption{This shows how the selection in \subfigref[(a)]{4a}, 
when copied and pasted into a text file, recovers the \LaTeX\ source \subfigref[(b)]{4b} 
that was used to specify the visual appearance of the mathematical content. 
 In \subfigref[(c)]{4c} 
 we see structure within a \pdt{/Formula} node, (see also Fig. \subfigref{5b})
 with leaf-nodes of  \pdt{/accesstag} structure nodes
 being \emph{marked content} of type \pdt{/AccessTag}. %; see Fig. \subfigref[b]{5a}. 
 This consists of a single space character 
 %. These `fake spaces' are tagged \pdt{/accesstag}marked content}, 
 carrying an \pdt{/ActualText} attribute which holds the replacement text; 
 as seen explicitly in the coding shown in Fig.\,\subfigref{5a}. %\ref{access-tags}a,b.
 The `fake spaces' are very narrow; when selected they can be seen very faintly in \subfigref[(c)]{4c} 
 within the ovals indicated, 
 at the outer edge of the the bounding rectangles of the outermost math symbols.}\label{access-select}
\end{figure}

\begin{figure}[htp]
%   Please do not adjust anything that affects vertical spacing within this Figure.
%   The subfigure captions have been carefully positioned over the large image.
\subfiglabel{5a}\vskip-\baselineskip
\vspace{-11pt}%
\setbox0=\hbox{\includegraphics[]{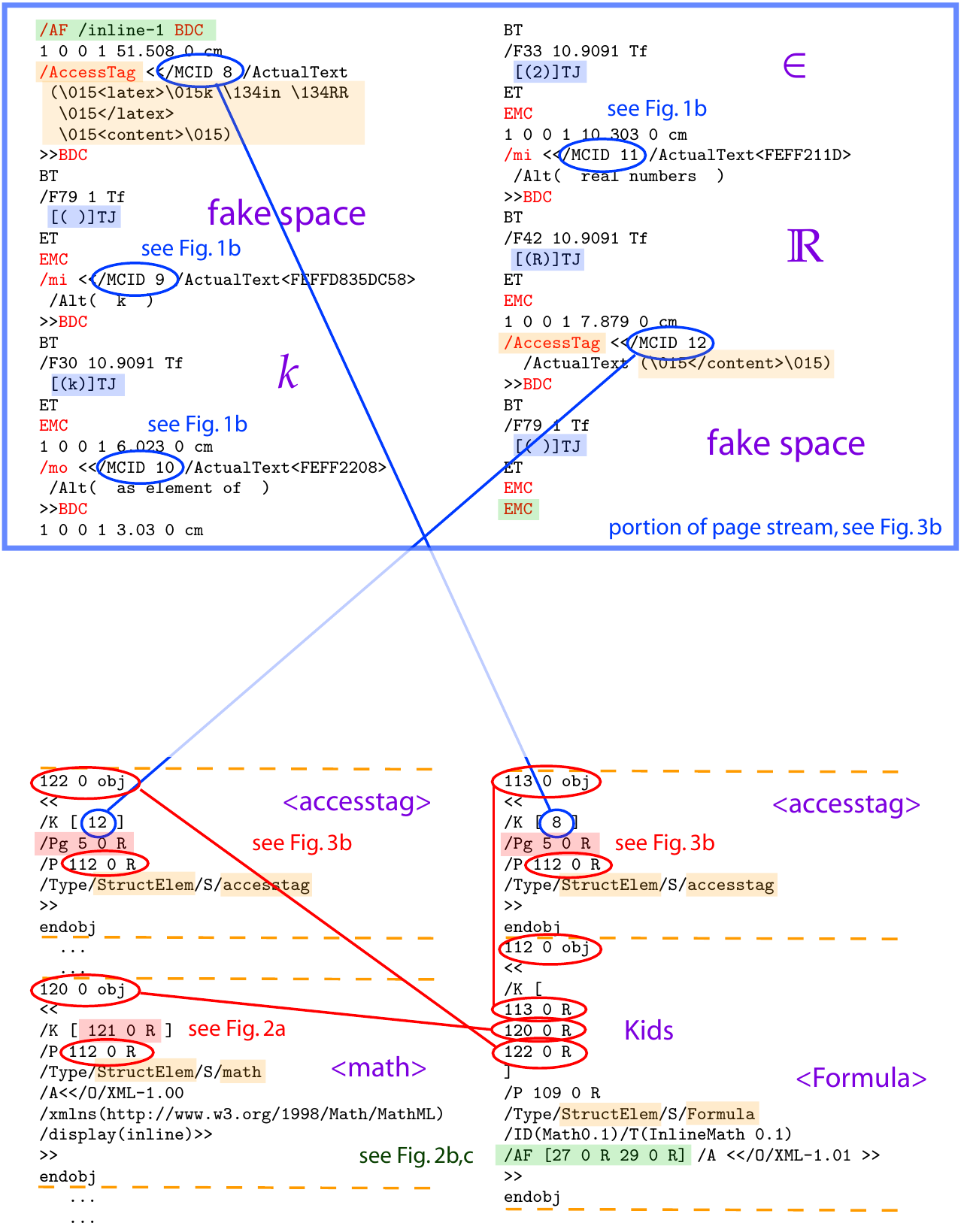}}%
\setbox0=\hbox{\kern-10pt\lower\ht0\box0}%
\ht0=0pt \dp0=0pt \box0\par
\setbox0=\vbox{\hsize=\linewidth
{\scriptsize
\begin{multicols}2
\begin{verbatim}
/AF /inline-1 BDC
1 0 0 1 51.508 0 cm
/AccessTag <</MCID 8 /ActualText 
 (\015<latex>\015k \134in \134RR
  \015</latex>
  \015<content>\015)
>>BDC
BT
/F79 1 Tf
 [( )]TJ
ET
EMC
/mi <</MCID 9 /ActualText<FEFFD835DC58>
 /Alt(  k  )
>>BDC
BT
/F30 10.9091 Tf
 [(k)]TJ
ET
EMC
1 0 0 1 6.023 0 cm
/mo <</MCID 10 /ActualText<FEFF2208>
 /Alt(  as element of  )
>>BDC
1 0 0 1 3.03 0 cm
BT
/F33 10.9091 Tf
 [(2)]TJ
ET
EMC
1 0 0 1 10.303 0 cm
/mi <</MCID 11 /ActualText<FEFF211D>
 /Alt(  real numbers  )
>>BDC
BT
/F42 10.9091 Tf
 [(R)]TJ
ET
EMC
1 0 0 1 7.879 0 cm
/AccessTag <</MCID 12
  /ActualText (\015</content>\015)
>>BDC
BT
/F79 1 Tf
 [( )]TJ
ET
EMC
EMC
\end{verbatim}%
\end{multicols}}%
%\removelastskip
}% end of box0
\vrule height \ht0 width 0pt \par
\removelastskip
%\vskip-.5\baselineskip
\leavevmode
\quad (a) 
Complete portion of the content stream corresponding to the mathematics shown
as selected in Fig.\,\subfigref{4a}. %\ref{access-select}(a). 
This is the same content as in Figures\,\subfigref{1b} and  \subfigref{3a} %\ref{formula-code}(b), 
but with the `\,...' parts there 
now showing the \pdt{/AccessTag} coding of a `fake space' with \pdt{/ActualText} attribute.
The \pdt{/AF ... BDC ... EMC} wrapping of Fig.\,\subfigref{3a} %\ref{assoc-cont}(a) 
is also shown.
Being part of the document's content, these space characters are also assigned \pdt{/MCID} numbers 
to be linked to structure nodes, as in \subfigref[(b)]{5b} below.\par
\subfiglabel{5b}\vskip-\baselineskip
\setbox0=\vbox{\hsize=\linewidth
{\scriptsize
\begin{multicols}2
\begin{verbatim}
122 0 obj
<<
/K [ 12 ]
/Pg 5 0 R
/P 112 0 R
/Type/StructElem/S/accesstag
>>
endobj
  ...
  ...
120 0 obj
<<
/K [ 121 0 R ]
/P 112 0 R
/Type/StructElem/S/math 
/A<</O/XML-1.00
/xmlns(http://www.w3.org/1998/Math/MathML)
/display(inline)>>
>>
endobj
   ...
   ...
113 0 obj
<<
/K [ 8 ]
/Pg 5 0 R
/P 112 0 R
/Type/StructElem/S/accesstag
>>
endobj
112 0 obj
<<
/K [
113 0 R
120 0 R
122 0 R
]
/P 109 0 R
/Type/StructElem/S/Formula 
/ID(Math0.1)/T(InlineMath 0.1)
/AF [27 0 R 29 0 R] /A <</O/XML-1.01 >>
>>
endobj
\end{verbatim}
\end{multicols}}%
\removelastskip
}% end of box0
\vrule height \ht0 width 0pt \par
\leavevmode
\quad (b) 
Portion of the structure tree as in Fig.\,\subfigref{1a}, %\ref{formula-code}a, 
but now showing how the `fake spaces' can be linked to structure nodes, here \pdt{/accesstag}.
The missing portions of Fig.\,\subfigref{1a}, %\ref{formula-code}a, 
indicated there by `\,...' are now filled-in,
but leaving out other parts whose purpose has already been explained.
Fig.\,\subfigref{4c}, %\ref{access-select}c 
shows the tagging opened out within the `\textsf{Tags}' navigation panel,
%but where 
with the \pdt{/accesstag} structure nodes selected.
\par
\vspace{-6pt}%
\caption{File content included as \pdt{/ActualText} for a `fake space', 
which itself can be tagged as \emph{marked content} linked to an \pdt{/accesstag} structure node.}\label{access-tags}%
\end{figure}

\section{Access-tags: attaching \LaTeX\ source to `fake' spaces.}
\label{sect3}%
\noindent
A third method allows inclusion of the \LaTeX\ source of mathematics
% or other technical material, within the body of a PDF document 
so that it may be readily extracted,
%This uses 
using just the usual \textsf{Select}/\textsf{Copy}/\textsf{Paste} actions. % to extract the information,
%so should work with a lot of 
This works with some existing PDF reader applications, 
including the freely available `Adobe Reader'.
It is achieved by making use of the \pdt{/ActualText} attribute \cite[\S 14.9.4]{ISO32000}
for a piece of \emph{marked content}, whether or not structure tagging is present.
%Thus this method 
It can be done by existing PDF-writing software that supports tagging of content, 
as %described
 in Sect.\,\ref{PDFtagging},
and specification of a value for the \pdt{/ActualText} attribute.

Fig.\,\ref{access-select} 
shows how this works, by tagging a `fake space' character immediately 
before mathematical content, and another immediately afterwards.
By selecting (see Fig.\,\subfigref{4a}) %\ref{access-select}a) 
then \textsf{Copy}/\textsf{Paste} the content into text-editing software, 
the result should be similar to %what is shown in 
Fig.\,\subfigref{4b}. %\ref{access-select}b.
The PDF reader must recognise%presence of 
\footnote{%
Adobe's `Reader' and `Acrobat Pro' certainly do, along with other software applications, 
but Apple's standard PDF viewers currently do not support \pdt{/ActualText}.}
 \pdt{/ActualText} and replace the copied content (e.g. a single font character) with its value.
%
%The portion of the \PDF\ page content stream corresponding to the mathematics shown as selected in Fig.\,\ref{access-select}.
The \emph{PDF\,name} \pdt{/AccessTag} tags a single `space' character \texttt{[( )]Tf} as \emph{marked content}, 
having replacement text in the \pdt{/ActualText} attribute; see Fig.\,\subfigref{5a}. %\ref{access-tags}a.
An \pdt{/MCID} number is not needed for this technique to work; 
these \emph{are} included in the example document\Exfootmark\ which is fully tagged\footnote{%
In PDF\,2.0 this will also need an association of \pdt{/accesstag} to \pdt{/Custom} within the \pdt{/RoleMap}
dictionary. The \pdt{/AccessTag} \emph{PDF\,name} can be replaced by \pdt{/Span}.}. % for Accessibility purposes.
%Similarly \pdt{/Alt} and \pdt{/ActualText} replacements are not needed on mathematical symbols;
%but are included to allow a reasonable audio presentation and text-extraction.

We refer to these tagged `fake spaces' as `Access-tags', since a motivation for their use is to allow 
Assistive Technology (e.g., a Braille reader) access to the \LaTeX\ source of mathematical content.
The spaces are `fake' in the sense that they are just 1pt in height and have nearly zero width.
This makes them hard to select by themselves, 
but nearly impossible to separate from the mathematical content which they accompany; 
see Fig.\,\subfigref{4c}. %\ref{access-select}c. 
They act as ordinary spaces when copied, but this is substituted with the \pdt{/ActualText} replacement,
if supported.
Another aspect of their `fakeness' is that they take no part in the typesetting, when using pdf-\LaTeX\ (post-2014).
Of course the same idea could be implemented in different ways with different PDF-producing software.

{\edef\0{\string\0}\edef\1{\string\1}%
One places into the initial `Access-tag' text of the \LaTeX\ source
 --- with care given to encode special characters% 
%`backslash', parentheses and line-ends
\footnote{%
Octal codes: \texttt{\string\134} for backslash, \texttt{\050} for `(' and  \texttt{\051} for `)', \texttt{\015} for line-end.}
% or `carriage-return'.}%
--- as the value of its \pdt{/ActualText} attribute.
The source coding is preceded by the string \verb|<latex>| and followed by \verb|</latex>| and \verb|<content>|,
with return-characters (in octal \texttt{\015}) to allow these `delimiters' to ultimately copy onto lines by themselves.
The trailing `Access-tag' just takes  \verb|</content>|. 
As a result, the eventual \textsf{Paste} gives text as shown in Fig.\,\subfigref{4b}. %\ref{access-select}b.
}% end of \edef\0  ... 

%\subsection{Accessible Text}
Assistive Technology (e.g., a Braille reader) % generally 
works either by
\begin{inparaenum}[(a)]
\item 
emulating \textsf{Copy}/\textsf{Paste} of on-screen portions of the document's window; or
\item
by directly accessing the `Accessible Text' view of the PDF document's contents%
\end{inparaenum}.
In both cases the \pdt{/ActualText} replacements are extracted.
(The `Accessible Text' view can be exported directly using Adobe's `Acrobat Pro' software, see \cite{ExPDFUA}.)
For mathematical symbols, where Figures~\subfigref{1b}, %\ref{formula-code}b 
\subfigref{3a} and \subfigref{5a} %\ref{access-tags}a
show the presence of both \pdt{/Alt} and \pdt{/ActualText} attributes, 
then the \pdt{/Alt} contributes to the `Accessible Text', 
whereas \pdt{/ActualText} supplies what is copied to the Clipboard for \textsf{Copy}/\textsf{Paste}.
In either case, a human reader when encountering \verb|<latex>| on a line by itself can choose whether
to read (or listen to) the following lines of \LaTeX\ coding, 
else use a \textsf{Find} action to skip down to where the next \verb|</latex>| occurs.
This is followed by a line containing \verb|<content>|. 
Similarly the human can read/listen % the subsequent content 
or skip down to where \verb|</content>| occurs.

%\section{Conclusion}
%This paper demonstrates three different ways in which the \LaTeX\ and \MathML\ source coding for mathematical content
%can be included within a document conforming to the \PDF/A-3u standard \cite{pdfA3}.
%When the document is also fully tagged, as in the particular example document\footnote{%
%This document should have come included with a \PDF\ version of this paper.}, 
%then the `Accessibility' requirements of the \PDF/UA standard \cite{PDF-UA1} can also be satisfied.
%It is hoped that some, if not all, of these methods will become standard in the mathematical and technical
%publishing industry within the not-too-distant future.

\paragraph*{Acknowledgements\/\upshape{:}}
The author wishes to acknowledge James Davenport and Emma Cliffe, at the University of Bath,
for valuable discussions regarding the needs of mathematics students having a visual disability.
The `fake space' idea emerged as a result. % discussions.
Thanks also to reviewers for suggesting, among other ideas, inclusion of the PDF language overview
as in Sect.\,\ref{pdf-overview},
%which 
allowing %ed the 
later sections to be written more succinctly.
Finally, I wish to thank members of  ISO TC 171 %Technical Committee 171, 
for encouragement and support %, especially
regarding `fake spaces' and other aspects.
\begin{quotation}\noindent\scriptsize
``The committee is really happy to have someone actually implementing math accessibility in PDF \dots '' \cite{Soiffer}
%and I am personally very grateful to see it happen and get your input.
\hfill --- Neil Soiffer, Senior Scientist, Design Science Inc., 12 April 2014.\par
\end{quotation}
\removelastskip
\goodbreak\goodbreak\goodbreak
\rightline{\attachfile[ucfilespec=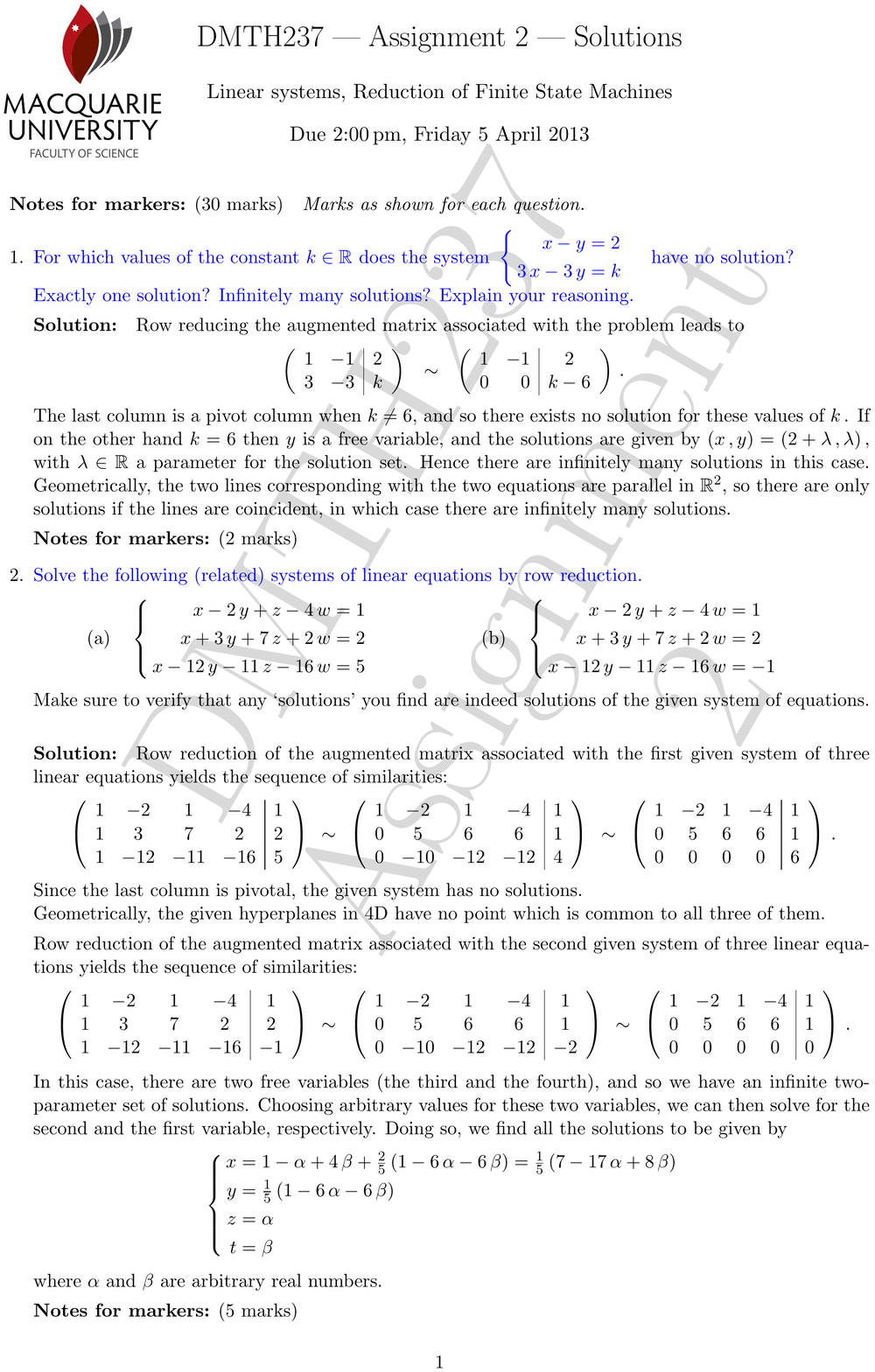,mimetype=application/pdf]{2013-Assign2-soln-3u.pdf}}%
\makeatletter
 \edef\thisembedfile{\csname atfi@fsobj@2013-Assign2-soln-3u.pdf\endcsname}%
 \addtodocassocfiles{\thisembedfile}%
\makeatother
\vspace{-\baselineskip}%
\vspace{-\baselineskip}%
\nobreak

\end{document}